Notice to the editor (not to be published): This manuscript has been authored by UT-Battelle, LLC, under Contract No. DE-AC05-00OR22725 with the U.S. Department of Energy. The United States Government retains and the publisher, by accepting the article for publication, acknowledges that the United States Government retains a non-exclusive, paid-up, irrevocable, world-wide license to publish or reproduce the published form of this manuscript, or allow others to do so, for United States Government purposes. The Department of Energy will provide public access to these results of federally sponsored research in accordance with the DOE Public Access Plan (http://energy.gov/downloads/doe-public-access-plan).

# Reconstruction of effective potential from statistical analysis of dynamic trajectories

A. Yousefzadi Nobakht,[1,*] O. Dyck,[2] D. B. Lingerfelt,[2] F. Bao,[3] M. Ziatdinov,[2] A. Maksov,[2,4] B.G. Sumpter,[2] R. Archibald,[5] S. Jesse,[2] S.V. Kalinin[2,†] and K.J.H. Law[5,6,‡]

[1] Department of Mechanical, Aerospace, and Biomedical Engineering, The University of Tennessee, Knoxville, TN 37996

[2] The Center for Nanophase Materials Sciences, Oak Ridge National Laboratory, Oak Ridge, TN 37831

[3] Department of Mathematics, Florida State University, Tallahassee, FL, 32304

[4] Bredesen Center for Interdisciplinary Research and Education, The University of Tennessee, Knoxville, TN 37996

[5] Computer Science and Mathematics Division, Oak Ridge National Laboratory, Oak Ridge, TN 37831

[6] School of Mathematics, University of Manchester, Manchester, UK

The broad incorporation of microscopic methods is yielding a wealth of information on atomic and mesoscale dynamics of individual atoms, molecules, and particles on surfaces and in open volumes. Analysis of such data necessitates statistical frameworks to convert observed dynamic behaviors to effective properties of materials. Here we develop a method for stochastic reconstruction of effective acting potentials from observed trajectories. Using the Silicon vacancy defect in graphene as a model, we develop a statistical framework to reconstruct the free energy landscape from calculated atomic displacements.

---

[*] ayousefz@vols.utk.edu

[†] Sergei2@ornl.gov

[‡] kodylaw@gmail.com

I. INTRODUCTION:

Molecular Dynamics (MD) simulation has emerged as one of the primary computational tools giving insight into the atomic scale processes in condensed matter physics, chemistry, and materials science. MD is a method for computationally evaluating the thermodynamic and transport properties of materials by solving the classical equations of motion at the molecular level. Classical Molecular Dynamics simulations provide the distribution of atoms and molecules in a material. They do not provide the distribution of electrons, in contrast to quantum mechanical calculations. MD simulation relies on the fact that the electron mass is much smaller than the mass of the nuclei and describes the time evolution of nuclear degrees of freedom using the classical Newtonian mechanics, whereas the influence of the electrons are captured by effective interatomic potentials[1]. MD method can provide wealth of information on material dynamics, single atom dynamics, and other equilibrium and non-equilibrium atomistic phenomenon [2-7].

However, one of the characteristic aspects of MD is the large amount of information it yields, necessitating the compression of the full set of time-dependent positions and conjugate force fields into a small number of physically relevant reduced variables. The complexity of many-particle systems is frequently the motivation for introducing mean-field approximations in order to simplify the descriptions given by complicated equations of motion. For example, linear response of an equilibrium system to a small external influence can be derived using the classical Kubo formula [8]. This approach can be used e.g. to calculate diffusion coefficients by space–time integral of the flux–flux equilibrium correlation function [9]. Other classical examples include order parameters in physics [10], slow variables in proteins [11], and the whole concept of reaction paths and reaction coordinates [12,13]. Until now, most of these information compression methods analyzed the full dynamics of the system parametrized as the time evolution of particle coordinates and conjugate variables (e.g. velocities).

At the same time, development of electron and scanning probe microscopy techniques have enabled detailed observations of atomic motion. The high spatial resolution afforded by modern instruments have enabled capturing the random walk of a point defect,[14] dopant diffusion in crystals,[15,16] as well as single atom catalytic etching[17,18] and growth[19]. Movement of single atom dopants in graphene [20-23] and the dynamics of Si clusters have also been observed [24-26]. Recent progress in automated analytical techniques leveraging machine learning have provided a path forward for detailed analysis of large data sets, extracting and tracking features of interest for a more comprehensive and statistically robust footing with regard to atomic dynamics.[27-29]. Analysis of such systems necessitates development of pathways to compress the MD simulation results to reconstruct effective potentials acting on the atoms, using collective dynamics of the atomic neighborhood in the analysis.

Here we develop a method for stochastic reconstruction of effective acting potentials from observed trajectories. For this, we perform the molecular dynamic simulation of the silicon vacancy defect in graphene for a known time-dependent excitation force and reconstruct the *effective* autonomous (time-independent) free energy landscape from calculated atomic displacements.

## II. MOLECULAR DYNAMICS SIMULATION:

The MD simulations were performed to predict the response of silicon vacancy defect in graphene to a known force applied to a single dopant Si atom embedded in a monovacancy in a graphene lattice. Periodic boundary conditions are applied in the all directions. To avoid interactions between periodic cells in the *z* direction, a large vacuum (50 nm) region separates both sides of the graphene sheet in this direction. Here, *x* and *y* are in-plane directions and *z* is out of plane direction. A 12×12 nm graphene sheet was used, with a single Si atom occupying a monovacancy in the graphene sample. (Fig. 1 a). Adaptive intermolecular reactive empirical bond order (AIREBO) [30] potential is used to model C-C bonded interactions, and the Tersoff potential [31] was employed for modeling Si-C interactions. The Large-scale Atomic/Molecular Massively Parallel Simulator (LAMMPS) [32] was used to carry out the molecular dynamics simulations; the total simulation time and time step size for all simulations are chosen to be 5.0 ns and 0.5 fs, respectively.

The initial structure was allowed to relax at the beginning of the simulation using an NPT (isothermal-isobaric) ensemble for 0.5 ns, which was followed by a NVT ensemble at a temperature of 300K for another 0.5 ns to achieve steady temperature (Temperature fluctuations < 3K) for entire system. After reaching a steady temperature for the system, a periodic force is applied, with the form $F = A(e^{0.5 \times m^2})$. Here *F* is the effective force applied to the Si atom in –*z* direction and *m* is a random generated number in the range [0, 1) and *A* is the factor defining the force magnitude. The force was applied as a short delta function with a duration of 100 fs and at a rate of every 2.5 ps. The time gap between each force application is long enough so the effect of the previous force is damped by the system. Fig. 1 b shows an example of the applied force distribution.

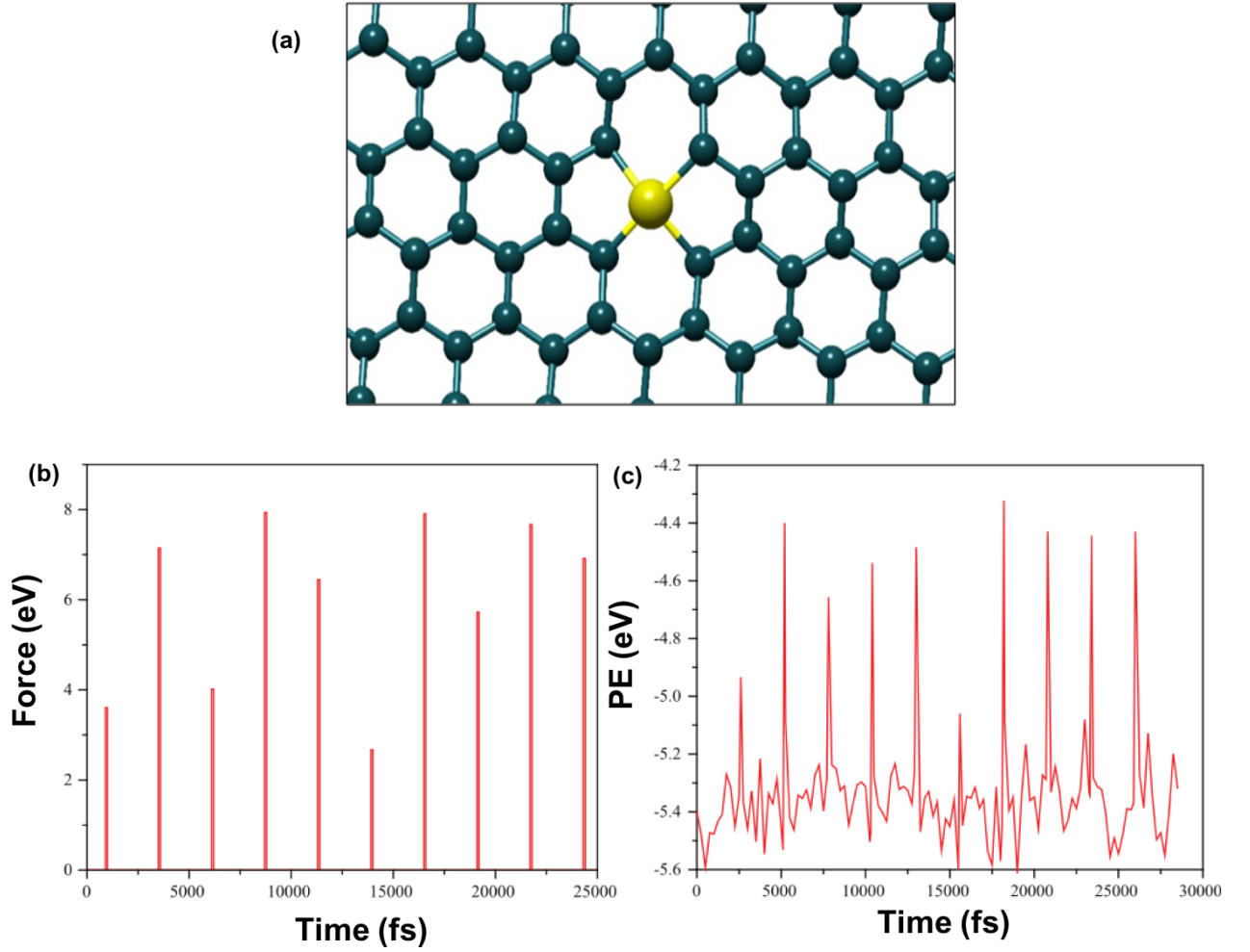

**Figure 1**. (a) Silicon atom configuration in graphene sheet. (b) Force distribution applied to Si atom. (c) Potential energy of Si atom with respect to time.

### III. RESULTS AND DISCUSSION:

In order to calculate potential, the position and potential energy of the Si atom was exported every time step during the NVE ensemble. The potential energy variation with respect to time for the Si atom under the influence of the force shown in Figure 1 b) is illustrated in Figure 1 c). To reconstruct the potential over the $(x, y, z)$ space, we choose the model potential consistent with site symmetry as $V(x, y, z; \theta) = V_\perp(x, y) + V_z(z)$, where the following parametric ansatz are employed for the individual terms

$$V_\perp(x,y) = a(x^2 + y^2)[1 + b\cos(4\tan^{-1}(x/y))], \quad V_z(z) = cz^2 + dz, \qquad (1\ a,b)$$

where $\theta = (a,b,c,d)$ are parameters. Note that in MD simulation all $(x,y,z)$ coordinates are available during the modeling, unlike the STEM experiment which is sensitive only to $(x,y)$.

For femto-scale dynamics of the MD simulation, on the order of the decorrelation time of the underlying process, it makes sense to utilize dynamical information in the reconstruction.

We leverage the sequential nature of the data and the dynamical information by using a sequential Monte Carlo sampling method to process the observed atomic trajectories and perform parameter estimation. First the energy, $E$, is fit to the quadratic variation of the process. As mentioned above, it is assumed that the potential $V$ has a specific form and is governed by a set of parameters $\theta$. It is furthermore assumed that the data arise as exact observations from a discretized first order Langevin stochastic differential equation (SDE) in this potential,

$$\zeta dX_t = -\nabla V(X_t)dt + \sqrt{2\zeta E}dW_t, \qquad (2)$$

where $X_t$ is the state, $E = k_B T$ is the characteristic energy and carries the dimension of energy, $\zeta$ is a fluctuation-dissipation constant with units of mass/time, and $W_t$ is a standard Brownian motion and carries the dimension of time.[33] This equation has invariant measure $\exp(-V/E)/Z$, where Z is a normalizing constant.

In this context we recover a Bayesian posterior distribution with a simple tractable form, and we can then implement a sequential Monte Carlo sampling framework [34] by leveraging the sequential form of the posterior. We can further reduce the computational expense by introducing a pseudo-dynamics on the parameters in the spirit of [35] and implementing a standard particle filter. The former results are presented for reconstruction of the posterior potential V given the MD simulation data.

Here a somewhat non-standard statistical model is introduced which simplifies the problem of static parameter estimation for stochastic differential equations (SDEs). In particular, discretization error is ignored, and observations are assumed exact, which yields a tractable posterior distribution. A sequential Monte Carlo sampler is then implemented for inference from the proposed posterior distribution.

We consider the following statistical model

$$\gamma_n(d\theta) = \prod_{i=1}^{n} p_\delta(X_i|X_{i-1}, \theta)p_0(d\theta), \qquad (3)$$

where $p_0$ is a prior on some parameter $\theta \in \mathbb{R}^d$, and $p_\delta$ is a $\delta$ time-step Euler approximation of the following SDE

$$dX_t = -\nabla U(X_t; \theta)dt + \sqrt{2D}dW_t, \qquad (4)$$

where $D \in \mathbb{R}^{p \times p}$ is a known diffusion constant (this will be revisited below), $W_t$ is Brownian motion on $\mathbb{R}^p$, $X_t \in \mathbb{R}^p$, $U: \mathbb{R}^{p+d} \to \mathbb{R}^p$ is Lipschitz. That is

$$p_\delta(X_i|X_{i-1}, \theta) = \exp(-\frac{1}{4D\delta}|X_i - X_{i-1} + \nabla U(X_t; \theta)\delta|^2). \qquad (5)$$

Define the posterior density of $\theta|X_0, ..., X_n$ by $\eta_n(d\theta) = \gamma_n(d\theta)/\gamma_n(1)$. The objective is to sample $\theta^{(i)} \sim \eta_J$ for $i = 1, ..., N$, and approximate expectations for bounded functions $\varphi: \mathbb{R}^d \to \mathbb{R}$

$$\eta_n(\varphi) := \int_{\mathbb{R}^d} \varphi(\theta)\eta_n(d\theta) = \int_{\mathbb{R}^d} \varphi(\theta)\eta_n(\theta)d\theta, \tag{6}$$

by

$$\eta_n(\varphi) \approx \frac{1}{N}\sum_{i=1}^{N} \varphi(\theta^{(i)}). \tag{7}$$

We cannot sample from this distribution, but we can obtain a convergent estimator using sequential Monte Carlo samplers,[36] as described below.

First, we analyze the system for a priori known D. We define

$$G_n(\theta) = \gamma_{n+1}(\theta)/\gamma_n(\theta) = p_\delta(X_{n+1}|X_n, \theta). \tag{8}$$

Let $M_n$ denote an MCMC kernel such that

$$(\eta_n M_n)(d\theta) := \int_{\mathbb{R}^d} \eta_n(d\theta') M_n(\theta', d\theta) = \eta_n(d\theta). \tag{9}$$

Let $\theta_0^{(i)} \sim p_0$. For $n = 1, \ldots, J$, repeat the following steps for $i = 1, \ldots, N$:

- Define $\widetilde{w}_n^i := G_{n-1}(\theta_{n-1}^{(i)})$ and $w_n^i = \widetilde{w}_n^i / \sum_{j=1}^{N} \widetilde{w}_n^i$.
- Resample. e.g. select $I_n^i \sim \{w_n^1, \ldots, w_n^N\}$, and let $\widehat{\theta}_n^{(i)} = \theta_{n-1}^{(I_n^i)}$.
- Mutate. Draw $\theta_n^{(i)} \sim M_n(\widehat{\theta}_n^{(i)}, \cdot)$.

We further define $\eta_n^N(\varphi) := \frac{1}{N}\sum_{i=1}^{N} \varphi(\theta_n^{(i)})$. Under mild conditions it is well known that as $N \to \infty$, one has that $\eta_n^N(\varphi) \to \eta_n(\varphi)$ almost surely. Rates, central limit theorem, and large deviations estimates can also be obtained.[37]

In practice, we do not have access to $D$ and we must derive it from the data as well. This is done offline in a preprocessing step before implementing the above method. Observe that $X_{n+1} - X_n \sim N(-\delta\nabla U(X_t; \theta), 2D\delta)$. It is clear then that the drift term is higher order in $\delta$ when computing an approximation of the quadratic variation of the limiting SDE

$$\widehat{Q} := \frac{1}{J+1}\sum_{n=0}^{J}(X_{n+1} - X_n)(X_{n+1} - X_n)^T. \tag{10}$$

Indeed, this gives a very good approximation to $2D\delta$, and we define $\widehat{D} := \widehat{Q}/(2\delta)$. As $\delta \to 0$, one has that $\widehat{D} \to D$.

An alternative approach to parameter inference is to introduce a pseudo-dynamic on the parameter as $\theta_{n+1} \sim N(\theta_n, C_\theta)$, and then solve the filtering problem for $\theta_n | X_1, \ldots, X_n$,[35] using a standard particle filter. We find that this approach can provide a reasonable approximate reconstruction with appropriately chosen $C_\theta \propto \delta$.

Now the long MD trajectory is used to fit an effective Langevin dynamics.[33] It is assumed that in this small mass and long time regime it is reasonable to ignore inertial effects and employ a first order Langevin equation for fitting.[33] First (10) is employed to predict the diffusion constant from the quadratic variation. It is observed that the dynamics is smooth on a femtosecond timescale, and so a timestep of $\delta = 0.025$ ps is considered. An additional difficulty arises here due to the deterministic time-dependent forcing in the $z$ direction, denoted by $F(t) = (0,0, f(t))^T$, as well as the difference in effective diffusion coefficients in the $(x, y)$ and $z$ directions, denoted $D_\perp$ and $D_z$, respectively. The Langevin dynamics now take the form

$$dX_t = -(\nabla U(X_t) + F(t))dt + \sqrt{2D}dW_t, \quad (11)$$

where $X = (x, y, z)^T$, and for simplicity we assume $D = \text{diag}\{D_\perp, D_\perp, D_z\}$. In this equation, (2) has been multiplied through by $\zeta^{-1}$, the inverse of the fluctuation-dissipation coefficient, which is now a matrix, also assumed diagonal, $\zeta = \text{diag}\{\zeta_\perp, \zeta_\perp, \zeta_z\}$. With a known temperature, i.e. *scalar* value of $k_b T = E = \zeta D$, one can estimate the fluctuation-dissipation coefficient matrix $\zeta$ from (10) by $\hat\zeta = E \widehat{D}^{-1}$. Using the diagonal assumption, one simply has $\zeta_\perp = E/D_\perp$ and $\zeta_z = E/D_z$. Since T=300K we have E = 0.026 eV. In particular, the estimates using (10) are $D_\perp = 0.16$ Å²/ps and $D_z = 0.55$ Å²/ps. Since $E = 0.024$ eV, this gives $\zeta_\perp = 0.16$ eV ps/Å² and $\zeta_z = 0.047$ eV ps/Å².

Comparing (2) with (4), one can observe furthermore that $\nabla U(X_t) = \zeta^{-1}\nabla V(X_t)$. It is assumed for simplicity that the dynamics obey an invariant measure, despite the time-dependent periodic forcing F(t). The forcing is therefore set to zero in the inference procedure. It is furthermore assumed that the potential is separable as $U(x, y, z) = U_\perp(x, y) + U_z(z)$, where the following parametric ansatz are employed for the individual terms

$$U_\perp(x, y) = a(x^2 + y^2)\left[1 + b\cos\left(4\tan^{-1}\left(\frac{x}{y}\right)\right)\right], \quad U_z(z) = cz^2 + dz, \quad (12)$$

and so the invariant measure is proportional to $\exp(-(U_\perp/D_\perp + U_z/D_z))$.

The method is used to fit $(a, b, c, d)$. Note that degeneracy may occur if $b$ is allowed to change sign, and ill-posedness will result from $a, c < 0$. So, it is assumed that $a = e^{\theta_1}$, $b = e^{\theta_2}$, $c = e^{\theta_3}$ and $d = \theta_4$. A prior $\theta \sim N(0, C_0)$ is used, with $C_0 = \sigma^2 I$, and $\sigma = 2$. The parameters $a\zeta_\perp, c\zeta_z$ have units eV/(Å²), $d\zeta_z$ has units of eV/(Å), and $b$ is dimensionless. The resulting posterior after $n = 1141$ observations separated by $\delta = 0.025$ ps is presented in Figure 2. The posterior expectations on the parameters are given by $\mathbb{E}a = 13.7$, $\mathbb{E}b = 0.0125$, $\mathbb{E}c = 0.882$, and $\mathbb{E}d = 0.479$. It is interesting to observe that $d \approx -\frac{1}{T}\int_0^T f(t)dt$. In the end, the posterior mean potential in eV is given by $V_\perp = U_\perp \zeta_\perp$ and $V_z = U_z \zeta_z$, i.e.

$$V_\perp(x,y) = 2.06(x^2 + y^2)\left[1 + 0.0125\cos\left(4\tan^{-1}\left(\frac{x}{y}\right)\right)\right], \quad V_z(z) = 0.039z^2 + 0.021z.$$

(18)

These are plotted in Figures 3 a) and 3 b) with observation trajectory overlaid.

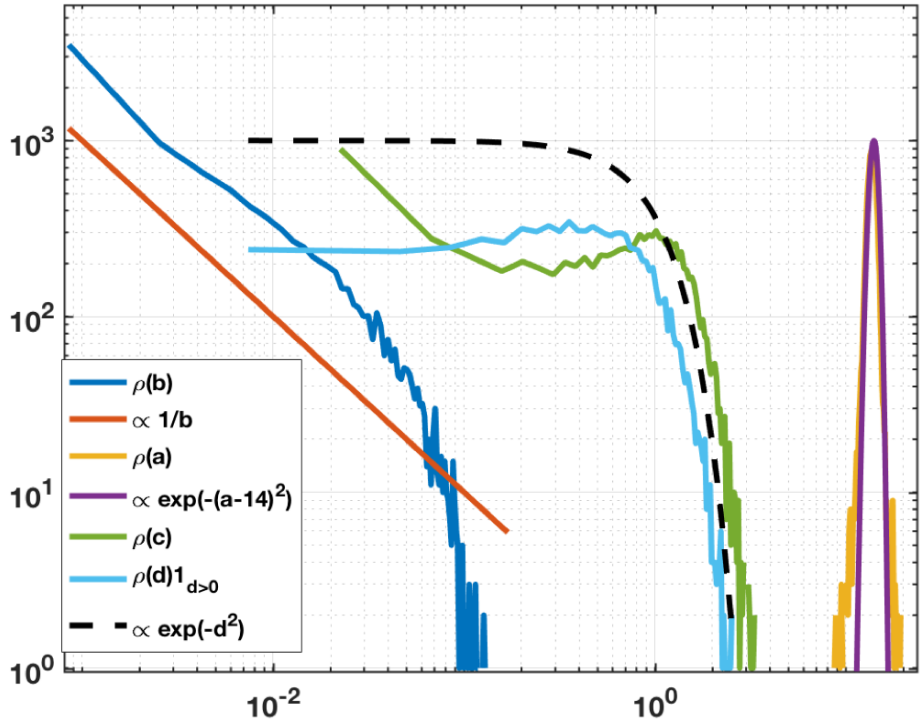

**Figure 2**: Marginal posterior densities of the pushforward of $\theta|X_1, \ldots, X_n$ to (a,b,c,d). The results are on a log scale and compared to some functional forms.

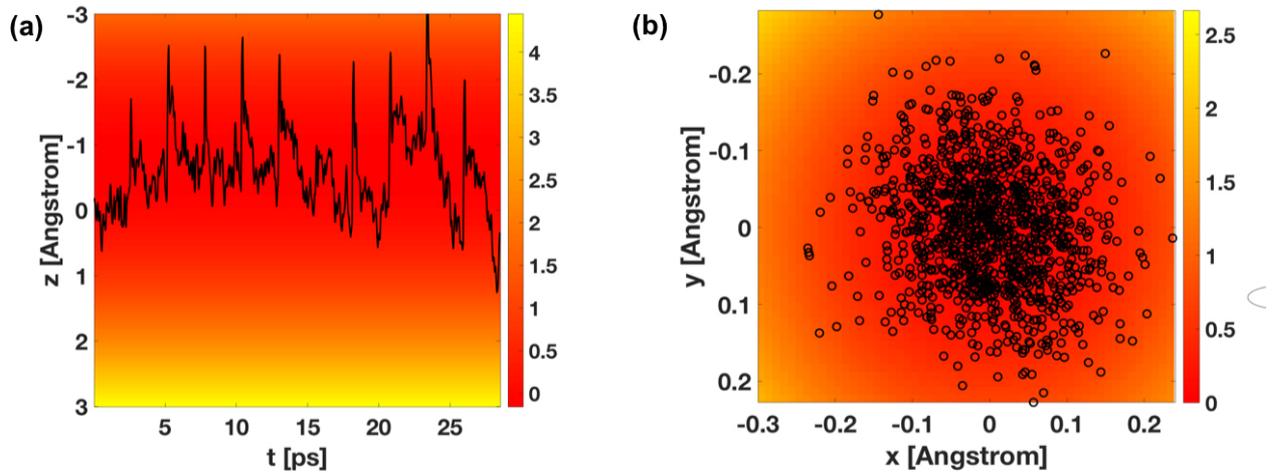

**Figure 3**: a) $V_z$ and trajectory of $z_t$. b) $V_\perp$ and trajectory of $(x_t, y_t)$. The relative energy scale (color) in (a,b) is in eV.

## IV. CONCLUSION:

To summarize, here we propose an approach to derive an effective single-particle potential from the observations of particle dynamics in multiple-particle systems. Here, the approach is developed for the specific case where observations of the full state are available, but it can be adapted for more complex cases as well with additional machinery, including partial and noisy observations. The latter is in fact the standard setting in the literature on hidden Markov models, where it is referred to as a state space model. This case is more challenging, as one must also infer (or marginalize over) the state as well.

**ACKNOWLEDGEMENTS:**

AYN would like to thank Dr. Seungha Shin and Jiaqi Wang for their expertise and assistance at Nano-Heat lab at the University of Tennessee. This work was also supported by The Alan Turing Institute under the EPSRC grant EP/N510129/1 (KJHL). This material is based upon work supported by the U.S. Department of Energy, Office of Science, Division of Materials Science and Engineering, Basic Energy Sciences (SVK, OD, SJ) and was performed at and partially supported by (MZ and BS) the Oak Ridge National Laboratory's Center for Nanophase Materials Sciences (CNMS), a U.S. Department of Energy, Office of Science User Facility. (RA, FB) would like to acknowledge the support by the Scientific Discovery through Advanced Computing (SciDAC) funded by U.S. Department of Energy, Office of Science, Advanced Scientific Computing Research through FASTMath Institutes and (FB) partial support by U.S. National Science Foundation under contract DMS-1720222.